\documentclass[british]{elsarticle}
\usepackage[T1]{fontenc}
\usepackage[latin9]{inputenc}
\usepackage{geometry}
\usepackage{parskip}
\usepackage{color}
\usepackage{verbatim}
\usepackage{float}
\usepackage{url}
\usepackage{varwidth}
\usepackage{amstext}
\usepackage{graphicx}

\makeatletter

\newcommand{\lyxmathsym}[1]{\ifmmode\begingroup\def\b@ld{bold}
  \text{\ifx\math@version\b@ld\bfseries\fi#1}\endgroup\else#1\fi}

\providecommand{\tabularnewline}{\\}
\newenvironment{cellvarwidth}[1][t]
    {\begin{varwidth}[#1]{\linewidth}}
    {\@finalstrut\@arstrutbox\end{varwidth}}

\makeatother

\usepackage{babel}
\begin{document}
\begin{frontmatter}
\title{Neutron Interaction Properties of Structural Materials for Multi-Grid
Neutron Detectors}
\author[ess]{A. Backis}
\author[ess]{C.-C. Lai}
\author[ess]{M. Aouane}
\author[ess]{P.P. Deen}
\author[ess]{K.G. Fissum}
\author[gla]{J.R.M. Annand\label{cauth}}
\ead{john.annand@glasgow.ac.uk}
\author[gla]{K. Livingston}
\author[ISIS]{D. Raspino}
\address[ess]{European Spallation Source ERIC, SE-221 00 Lund, Sweden}
\address[gla]{School of Physics and Astronomy, University of Glasgow G12 8QQ, Scotland,
UK}
\address[ISIS]{ISIS Facility, Rutherford Appleton Laboratory, Harwell Campus, Oxfordshire
OX11 0QX, UK}
\cortext[cauth]{Corresponding Author}
\begin{abstract}
The T-REX neutron time-of-flight spectrometer at the European Spallation
Source will use Multi-Grid Technology, which relies on thin ${\normalsize \mathrm{B_{4}C}}$
coatings on the Al blades of the grids to detect scattered thermal
neutrons. Following a Monte Carlo study of internal shielding to suppress
neutron multiple scattering in T-REX, the neutron transmission and
scattering properties of 12 shielding-material samples have been measured
at the ISIS spallation neutron source. Neutron transmission was measured
on the EMMA beam line at wavelengths 0.5-4.7~Å, using a 2D-position-sensitive,
neutron GEM detector, while neutron scattering was measured for 6
of the samples at the Merlin spectrometer, at wavelengths 0.72, 1.28,
1.85 and 2.41 Å. The present tests show that a ${\normalsize \mathrm{B_{4}C}}$/Al
composite material, plated with Ni to stop intrinsic alpha background,
is an effective neutron absorber, suitable for incorporation in the
Multi-Grid structures which detect the neutrons in inelastic neutron
spectrometers .
\end{abstract}
\end{frontmatter}

\section{Introduction}

\begin{figure}[H]
\includegraphics[width=1\columnwidth]{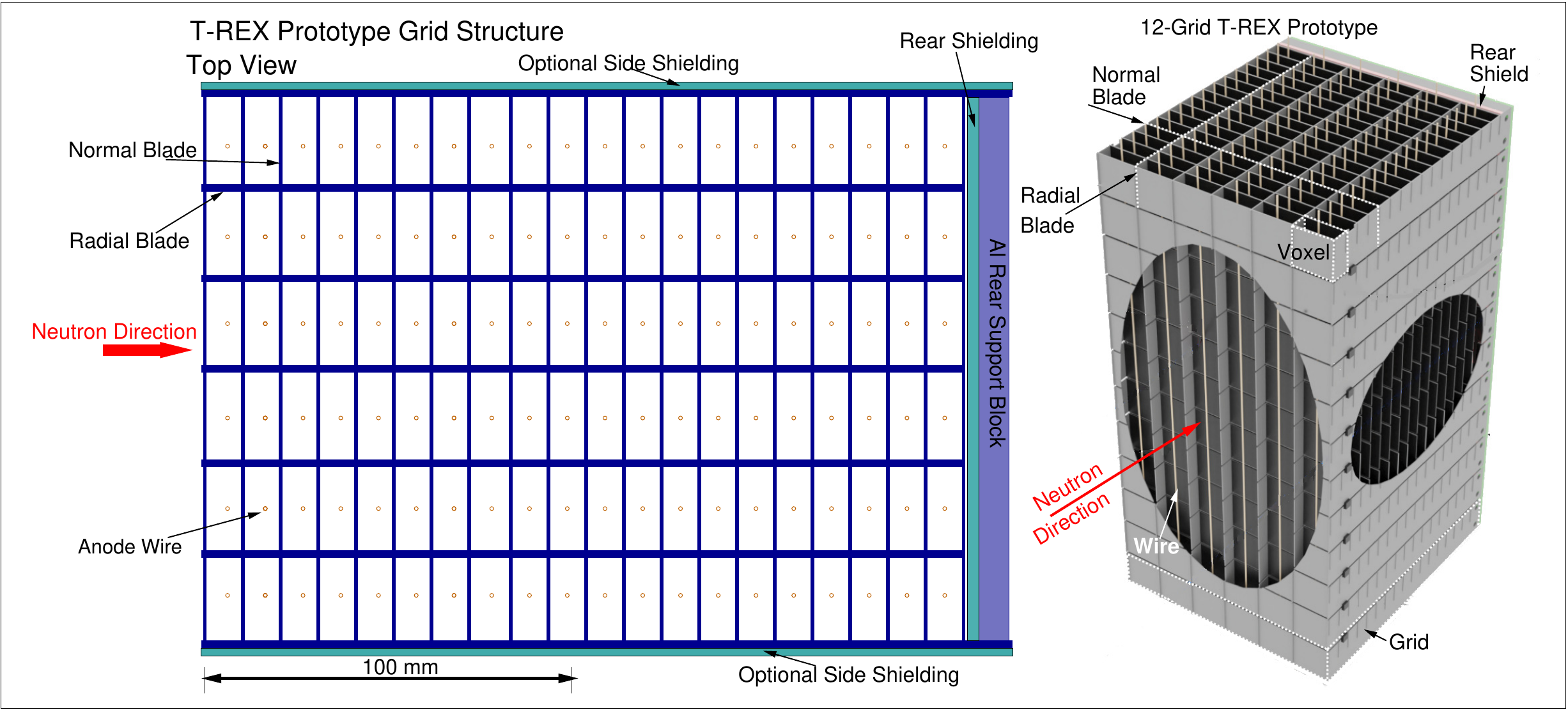}

\caption{\protect\label{fig:Multigrid}Left: prototype T-REX grid. Each voxel
has an internal dimension of 23.5~mm (x) by 24.0~mm (y) by 9.5~mm
(z). Right: 3D view of a 12-grid T-REX prototype, with outer sections
cut away to reveal the internal structure.}
\end{figure}

Multi-Grid (MG) technology for detection of sub-eV neutrons, originally
developed at ILL \citep{ILL}, will be used in the T-REX bispectral
chopper spectrometer which is under construction at the European Spallation
Source (ESS) \citep{ESS}. The Multi-Grids are stacks of grids, each
a rectangular lattice of normal and radial Al blades (Fig.\ref{fig:Multigrid}).
The grids form the cathodes of a voxelised proportional counter (VPC),
with wires strung through the centres of each grid voxel providing
the anodes. The normal blades are coated with $\mathrm{^{10}B}$-enriched
$\mathrm{B_{4}C}$, and neutron capture in the $\mathrm{B_{4}C}$
film produces $\mathrm{^{4}He}$ and $\mathrm{^{7}Li}$ ions, one
of which escapes into the VPC gas (e.g. Ar-CO2) giving a detectable
signal. Al is the main structural material for both the grids, their
support mechanics and the gas containment vessels, as its neutron
absorption cross section is relatively low. Nevertheless neutrons
can scatter internally in the Al and other nearby materials before
they eventually convert and this distorts the time-of-flight (TOF)
and angle-of-detection measurement, thereby distorting the energy
and momentum-transfer measurement by the spectrometer. 

Internal neutron scattering can be suppressed by installation of internal
shielding \citep{Shielding,Shield1,Shield2}, at the sides and rear
of the MG and on the radial blades of the grids. Monte Carlo simulations
of neutron scattering from Vanadium into T-REX \citep{G4-Trex} indicate
that internal shielding materials based on $\mathrm{B_{4}C}$ can
result in a major reduction in internal scattering within the spectrometer.
Scintered $\mathrm{B_{4}C}$ sheet would be a possibility for side
and rear shields, but structurally this is a difficult material and
it is unsuitable for radial blades, which must be electrically conductive.
A number of materials containing $\mathrm{B_{4}C}$ are available
and their shielding effectiveness has been investigated using the
Geant4 \citep{G4} model of T-REX. Subsequently neutron transmission
and scattering properties of good candidate shielding materials were
measured at the EMMA \citep{EMMA} and MERLIN \citep{Merlin} beam
lines of the neutron-spallation facility ISIS and compared to Geant4
simulations of these test measurements.

\section{\protect\label{sec:Neutron-Transmission-Tests}Neutron Transmission
Tests at EMMA}

\begin{figure}[H]
\includegraphics[width=1\columnwidth]{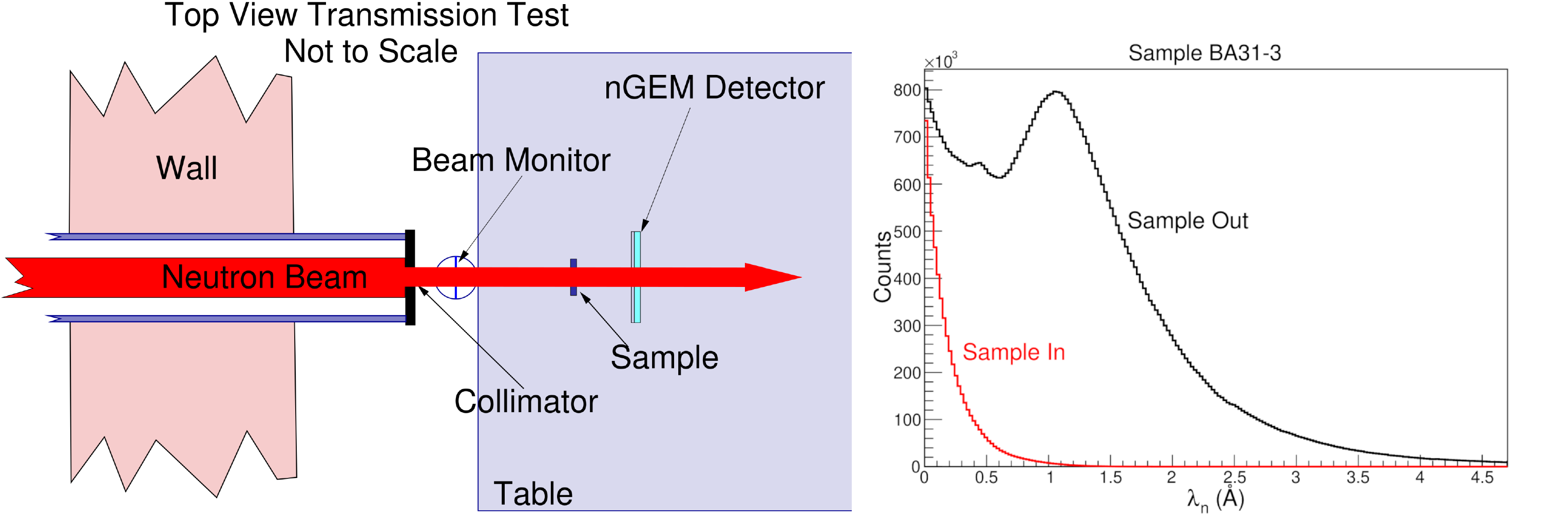}

\caption{\protect\label{fig:EMMA-white-neutron}Left: schematic diagram of
the EMMA neutron transmission test. Right: comparison of sample in/out
wavelength spectra derived from TOF at the nGEM detector. The sample
was BA31-3 (see Table~\ref{tab:Shielding-candidates}).}
\end{figure}

Measurements of neutron transmission through samples of shielding
material were performed on the EMMA beam line, with the Fermi chopper
powered down to produce a white neutron beam. A schematic of the experimental
setup is shown in Fig.~\ref{fig:EMMA-white-neutron} (left plot),
while the neutron wavelength spectrum is displayed in Fig.~\ref{fig:EMMA-white-neutron}
(right plot, sample out). The samples are listed in Table~\ref{tab:Shielding-candidates}.

The neutron beam was collimated by movable $\mathrm{B_{4}C}$ jaws,
set to produce a rectangular aperture of $25\times50\:\mathrm{mm^{2}}$.
It then passed through a low efficiency beam monitor (BM), the sample
under study, and a 2D, position-sensitive neutron gas electron multiplier
(nGEM) detector \citep{nGEM}. The BM was a 0.25~mm sheet of GS1
scintillating glass, with a neutron efficiency of $\sim0.5$\% at
a wavelength of $\sim1$~Å. The nGEM was a $10\times10\:\mathrm{mm^{2}}$~
GEM chamber read out by 128 x and 128 y strips (0.8~mm pitch), with
a thin layer of{\footnotesize{} $\mathrm{^{10}B}$ }deposited on the
cathode to convert neutrons to detectable ions. The intersections
between x and y strips define 16384 pixels. TOF was recorded for the
nGEM strips and BM. Samples were rectangular sheets of the material
under study inserted into the neutron beam, downstream of the BM and
perpendicular to the beam direction. 

\begin{table}[H]
{\scriptsize{}%
\begin{tabular}{|c|c|c|c|c|c|c|c|}
\hline 
{\small Name} & {\small Substrate} & {\small Coating} & {\small$\mathrm{B_{4}C}$(\%)} & {\small$\mathrm{^{10}B}$(\%)} & \begin{cellvarwidth}[t]
\centering
{\small Slope}{\small\par}

{\small$\pm5$\%}
\end{cellvarwidth} & {\small G4} & {\small$k$}\tabularnewline
\hline 
\hline 
{\small Al} & {\small HP Al 0.5~mm} & {\small --} & {\small --} & {\small --} & {\small 0.00} &  & {\small 0.00}\tabularnewline
\hline 
{\small Al-5} & {\small HP Al 0.5~mm} & {\small Sputtered $\mathrm{B_{4}C}$ 5~$\mu$m} & {\small 100} & {\small 97} & {\small 0.18} & {\small 0.23} & {\small 0.23}\tabularnewline
\hline 
{\small Dr-10} & {\small Al 0.5~mm} & {\small Drip$\mathrm{B_{4}C}$/Epoxy 10 $\mu$m} & {\small 73.8} & {\small 92} & {\small 0.47} & {\small 0.25} & {\small 0.25}\tabularnewline
\hline 
{\small Dr-20} & {\small Al 0.5~mm} & {\small Drip$\mathrm{B_{4}C}$/Epoxy 20 $\mu$m} & {\small 73.8} & {\small 92} & {\small 1.04} & {\small 0.50} & {\small 0.50}\tabularnewline
\hline 
{\small Sp1.4} & {\small Al 1~mm} & {\small Sprayed$\mathrm{B_{4}C}$/Al 300 $\mu$m} & {\small 52} & {\small 19.9} & {\small 0.56} & {\small 1.41} & {\small 1.39}\tabularnewline
\hline 
{\small Sp1.6} & {\small Al 1~mm} & {\small Sprayed$\mathrm{B_{4}C}$/Al 300 $\mu$m} & {\small 52} & {\small 19.9} & {\small 1.19} & {\small 1.41} & {\small 1.39}\tabularnewline
\hline 
{\small BA31-1} & {\small$\mathrm{Al/B_{4}C}$ 1 mm} & {\small --} & {\small 31} & {\small 19.9} & {\small 1.54} & {\small 1.52} & {\small 1.51}\tabularnewline
\hline 
{\small NiBA31-1} & {\small$\mathrm{Al/B_{4}C}$ 0.95 mm} & {\small Plated Ni 25 $\mu$m} & {\small 31} & {\small 19.9} & {\small 1.43} & {\small 1.45} & {\small 1.43}\tabularnewline
\hline 
{\small BA31-2} & {\small$\mathrm{Al/B_{4}C}$ 2 mm} & {\small --} & {\small 31} & {\small 19.9} & {\small 3.06} & {\small 3.03} & {\small 3.03}\tabularnewline
\hline 
{\small BA31-3} & {\small$\mathrm{Al/B_{4}C}$ 3 mm} & {\small --} & {\small 31} & {\small 19.9} & {\small 4.55} & {\small 4.59} & {\small 4.54}\tabularnewline
\hline 
{\small BA25-3} & {\small$\mathrm{Al/B_{4}C}$ 3 mm} & {\small --} & {\small 25} & {\small 19.9} & {\small 3.54} & {\small 3.67} & {\small 3.68}\tabularnewline
\hline 
{\small BA25-8} & {\small$\mathrm{Al/B_{4}C}$ 8 mm} & {\small --} & {\small 25} & {\small 19.9} & {\small 9.54} & {\small 9.93} & {\small 9.82}\tabularnewline
\hline 
\end{tabular}}{\scriptsize\par}

\caption{\protect\label{tab:Shielding-candidates}Shielding candidate materials.
Column `Substrate' is the substrate material of the sample where
HP stands for High Purity. Coating (if applied) is the coating applied
to the substrate. `$\mathrm{B_{4}C}(\%)$' is the content by weight
of $\mathrm{B_{4}C}$ in the substrate or coating and `$\mathrm{^{10}B}$(\%)'
is the $\mathrm{^{10}B}$ content (19.9\% in natural B). Slope is
the exponential decay constant fitted to the averaged $R_{io}$ transmission
measurement displayed in Fig.~\ref{fig:Average-ratio}. G4 is the
decay constant fitted to the Geant4 calculation and $k=2133N_{10}t$
is a hand calculation (see text) of the decay constant, based on a
$\mathrm{^{10}B}$ capture cross section \citep{10B} $\sigma=2133\lambda_{n}$~b.}
\end{table}

\begin{figure}[H]
\includegraphics[width=1\columnwidth]{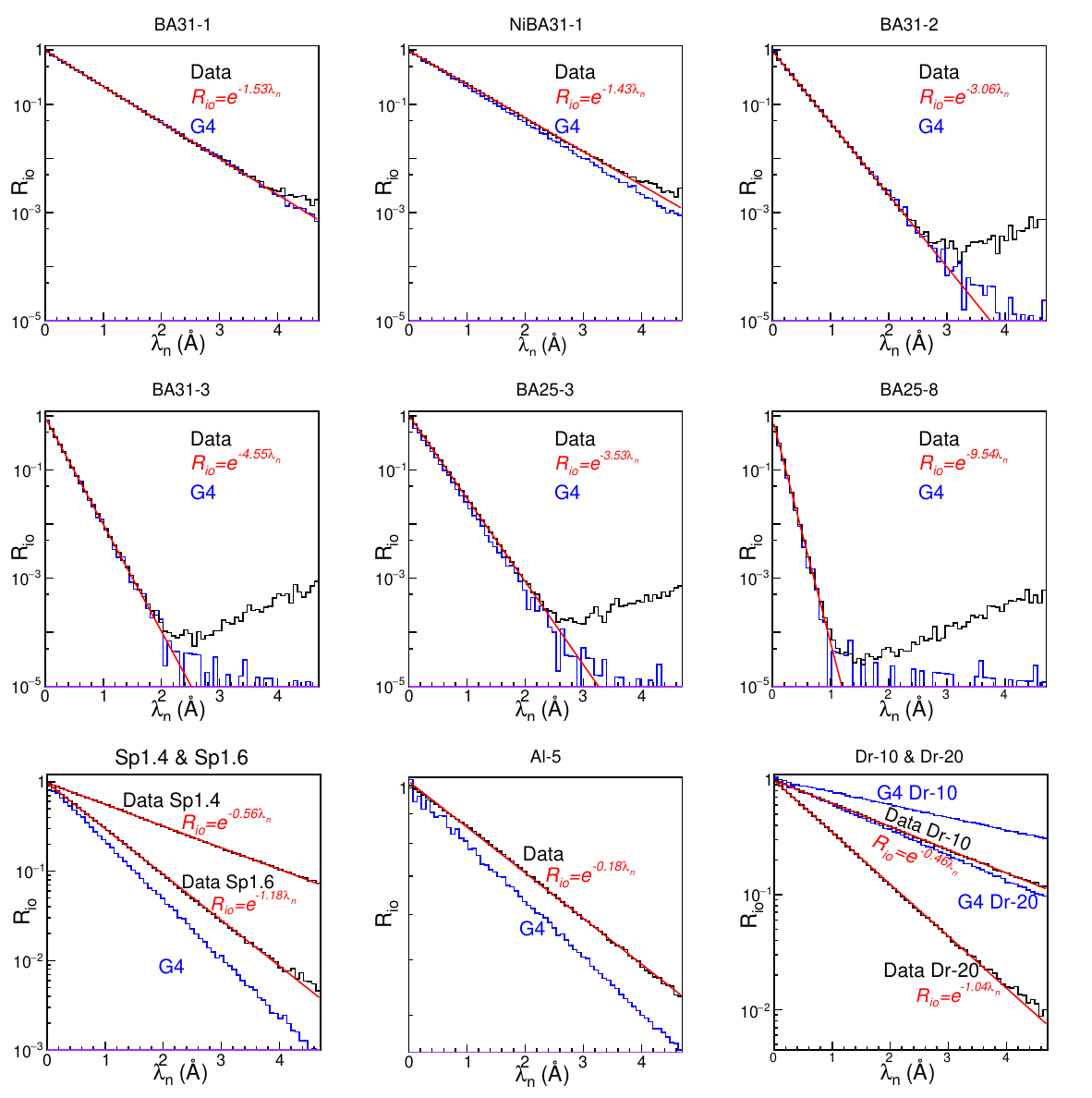}

\caption{\protect\label{fig:Average-ratio}Average ratio $R_{io}$. The data
are displayed in black and the red lines are exponential fits to the
$R_{io}$ data. Geant4-simulated distributions (G4) are shown in blue. }
\end{figure}

\begin{figure}[H]
\includegraphics[width=1\columnwidth]{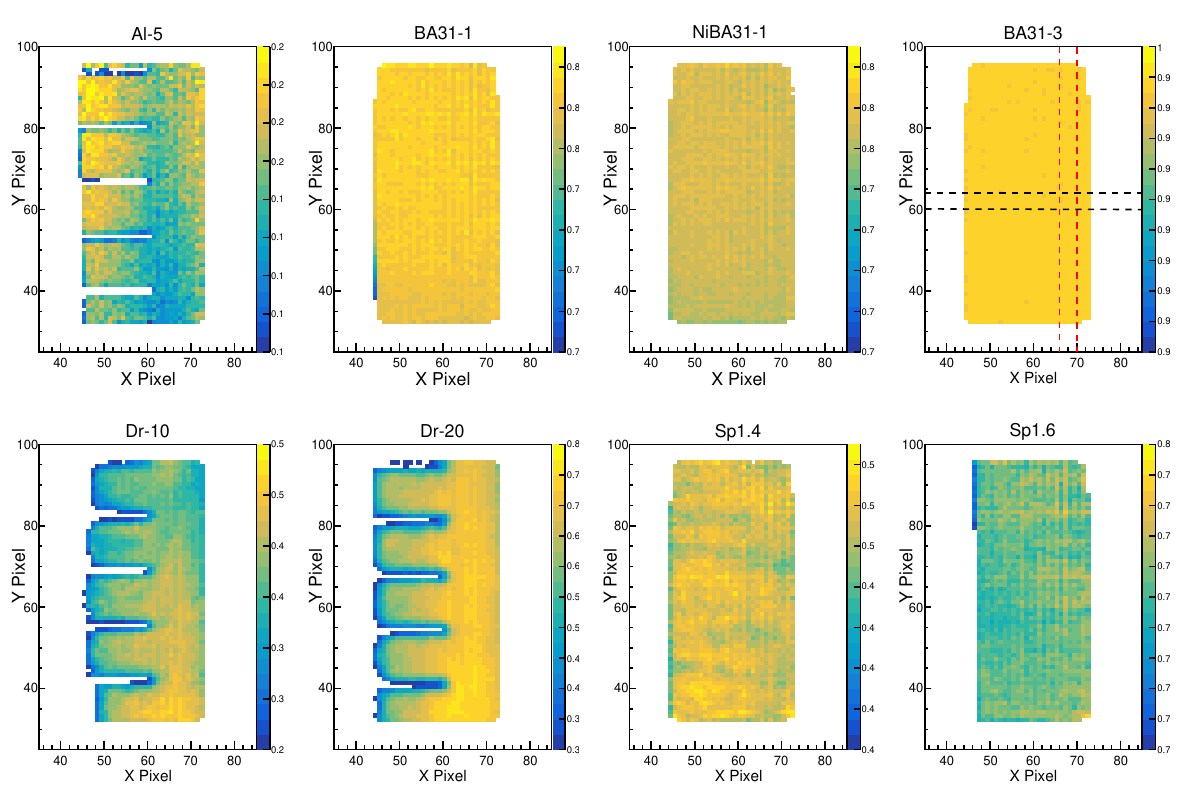}

\caption{\protect\label{fig:2D-images}2D images of neutron absorption $1-R_{io}$
on eight samples. Note that the vertical scales of the plots are different
to highlight any variations across the samples. Horizontal black dotted
lines (plot BA31-3) show the limits for the X-projections in Fig.~\ref{fig:1D-slices-of},
while the vertical red dotted lines show the Y-projection limits.}
\end{figure}

Runs were made with sample in and sample out, for equal accumulations
of proton charge on the spallation target. Counts in the nGEM pixels
were normalised to equal neutron counts in the beam monitor and the
ratio ($R_{io}$) of sample-in/sample-out counts gives a measure of
the neutron transmission probability through the sample. Fig.~\ref{fig:Average-ratio}
displays $R_{io}$, averaged over the area of the sample, as a function
of neutron wavelength $\lambda_{n}$, Those nGEM pixels at the edges
of the sample shadow and those pixels showing the slots in the sample
(see Fig.~\ref{fig:2D-images}) were omitted from the averaging procedure.
If neutron capture on $\mathrm{^{10}B}$ is the dominant determinant
of the transmission and the capture cross section depends linearly
on wavelength ($\sigma=2133\lambda_{n}$~b \citep{10B}), then $R_{io}=\exp(-k\lambda_{n})$.
Here $k=2133N_{10}t$ , where $N_{10}$ is the density of $\mathrm{^{10}B}$
atoms in the sample and $t$ is the sample thickness. Thus on a logarithmic
scale $R_{io}$ should appear linear in $\lambda_{n}$. In the higher
$k$ samples there is a significant departure from linearity at longer
wavelengths.

The EMMA beam line and end station are not evacuated so that there
is significant air scattering of neutrons and in addition, the region
around the collimator is weakly shielded. Thus the neutron beam has
a substantial halo which produced background counts in the nGEM, irrespective
of any sample in the path of the beam. At longer wavelengths, with
the more opaque samples, the background becomes significant with respect
to the sample-in counts and this distorts the ratio. The 480~keV
gamma produced in 94\% of $\mathrm{^{10}B}$ capture events can in
principle also add to background. As $k$ increases the number of
neutrons reaching the nGEM decreases, but the number of capture gammas
increases. However the Geant4 simulation, which has a simplified model
of the nGEM but does not include beam halo effects, predicts that
the gamma contribution to background will be very small. 

The measured and Geant4 simulated transmission curves in Fig.~\ref{fig:Average-ratio}
were fitted with exponentials and the resulting slope constants are
given in Table~\ref{tab:Shielding-candidates}. Longer wavelengths,
where the measured ratio departs significantly from exponential, were
excluded from the fits and the uncertainty in the fitted value is
$\sim\pm5$\%. The Geant4 values are consistent with the hand calculations
of $k$ described above. 

\begin{figure}[H]
\includegraphics[width=1\columnwidth]{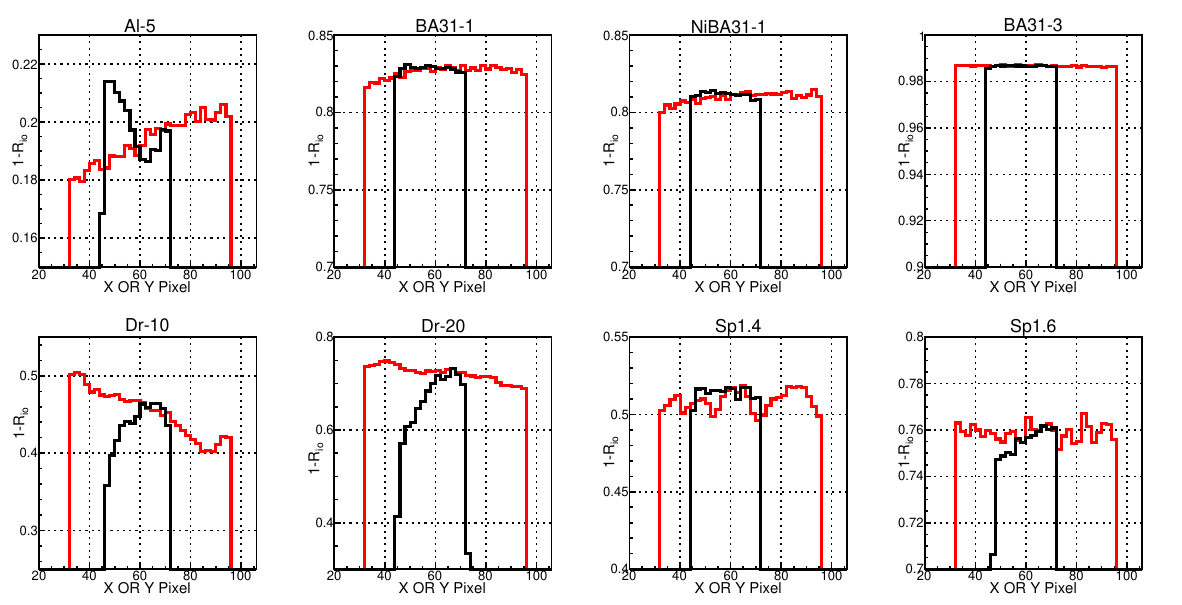}

\caption{\protect\label{fig:1D-slices-of}1D projections of the 2D neutron
absorption plots in Fig.~\ref{fig:2D-images}, with X-projections
(black) summed over y bins 60-64 and Y-projections (red) summed over
x bins 66-70. Note that the Y-axis scales are different as $1-R_{io}$
varies considerably from sample to sample.}

\end{figure}

Fig.~\ref{fig:2D-images} displays for eight samples the probability
of absorption ($1-R_{io}$), integrated over the wavelength range
$\lambda_{n}=0.5\lyxmathsym{\textendash}3.0$~Å, for nGEM pixels
seeing direct beam. Taking X- and Y- projections of the 2D plots gives
the 1D slices shown in Fig.~\ref{fig:1D-slices-of}. The X-projections
have been summed over Y pixels 60-64 and the Y-projections summed
over X pixels 66-70.

Samples Al-5, Dr-10 and Dr-20 have slots cut for assembly into a grid,
which appear in the 2D images. Al-5, Dr-10, Dr-20, Sp1.4 and Sp1.6
show variable, position-dependent absorption. Cold spray coatings
(Sp1.4, Sp1.6) display a faint horizontal banded structure in $1-R_{io}$,
which is also visible to the naked eye. Al-5 sputtered coating shows
relatively small variation in absorption across the face of the sheet,
while dripped coatings (Dr10, Dr20) show much larger variations. Al/$\mathrm{B_{4}C}$
composite (BA31-1, NiBA31-1, BA31-3) absorption is relatively uniform
across the samples.

The findings of the transmission tests are summarised as follows:
\begin{enumerate}
\item For the ${\normalsize \mathrm{Al/B_{4}C}}$ composite samples the
measured decay constants are close to $k$, so that the $\mathrm{^{10}B}$
concentration and thickness of these samples are close to the nominal
values given in Table~\ref{tab:Shielding-candidates}. Note that
the Geant4 calculation for NiBA31-1 was made for 1~mm of composite,
while the actual thickness was found to be $\approx0.95$~mm. The
G4 entry for NiBA31-1 in Table~\ref{tab:Shielding-candidates} has
been scaled by a factor 0.95. From Fig.~\ref{fig:2D-images}, \ref{fig:1D-slices-of}
the mix of ${\normalsize \mathrm{B_{4}C}}$ and Al appears quite homogeneous
throughout these samples
\item The Al-5 decay constant is around 78\% of $k$. Sputtering of ${\normalsize \mathrm{B_{4}C}}$
powder on Al does not produce a completely uniform film of coating
and it is likely that the average thickness of ${\normalsize \mathrm{B_{4}C}}$
is $\sim4\:\mu$m. The Al-5 plots in Fig.~\ref{fig:2D-images},\ref{fig:1D-slices-of}
show some smooth variation in $1-R_{io}$ at the $\sim10$\% level.
\item Cold spray coatings of a ${\normalsize \mathrm{Al/B_{4}C}}$ composite
on Al sheet, samples Sp1.4 and Sp1.6, both show decay constants considerably
smaller than $k$ calculated for a nominal 52\% ${\normalsize \mathrm{B_{4}C}}$
content and 300~$\mu$m coating thickness. Sp1.6 has a factor $\sim2$
more $\mathrm{^{10}B}$ in the coating than Sp1.4 and both samples
show a banded structure in the 2D plots of $1-R_{io}$ (Fig.~\ref{fig:2D-images}),
but the variations in $\mathrm{1-R_{io}}$ (Fig.~\ref{fig:1D-slices-of})
are at the few percent level .
\item Dripped coatings of ${\normalsize \mathrm{B_{4}C}}$/epoxy mix on
Al sheet give decay constants a factor $\sim2$ higher than the nominal
$k$. The nominal concentration of $\mathrm{^{10}B}$ is already 92\%,
so the coating thickness is likely greater than specified. The position
dependence of the thickness shows substantial variation in Fig.~\ref{fig:2D-images},
\ref{fig:1D-slices-of}.
\end{enumerate}

\section{\protect\label{subsec:Merlin-Measurement}Neutron Scattering Measurement
at Merlin}

\begin{figure}[H]
\begin{center}\includegraphics[width=0.5\columnwidth]{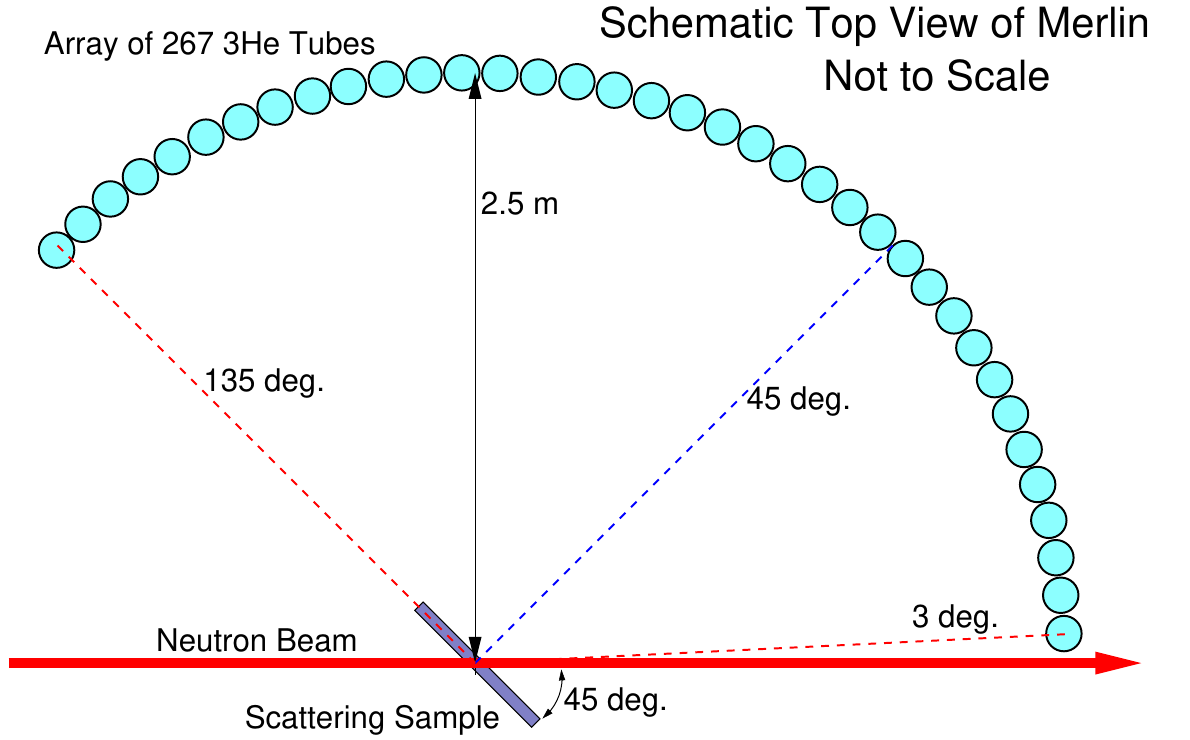}\end{center}

\caption{\protect\label{fig:Merlin-schematic}Simplified schematic diagram
of the sample-scattering test at Merlin.}

\end{figure}

Merlin at ISIS \citep{Merlin} is a TOF spectrometer with medium energy
resolution and a large solid angle ($\pi$~sr) spanning scattering
angles from 3-135$\lyxmathsym{\textdegree}$ in the horizontal plane.
The neutron detectors are $\mathrm{^{3}He}$ tubes 25~mm in diameter
and 3~m in length, situated on an arc 2.5~m from the scattering
sample.

\begin{figure}[H]
\includegraphics[width=1\columnwidth]{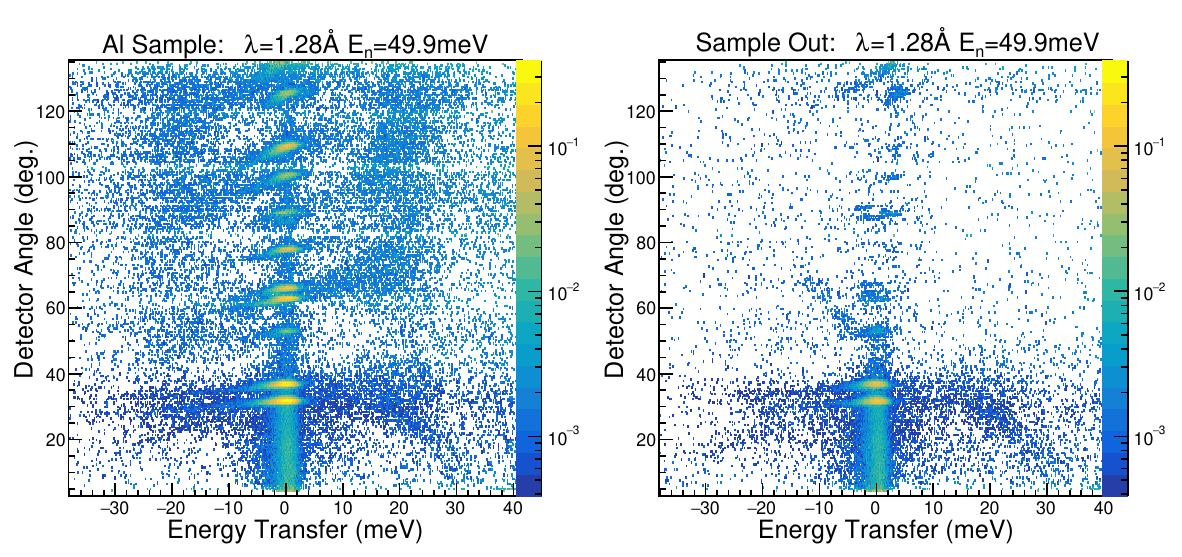}

\caption{\protect\label{fig:Merlin-2D}2D plots of polar angle vs. energy transfer
measured at Merlin. The logarithmic z scale is the same for both plots.}

\end{figure}

Samples Al, Al-5, BA31-1, NiBA31-1, BA25-3 and Dr-20 were placed at
the Merlin sample position, angled at 45$\lyxmathsym{\textdegree}$
with respect to the beam direction, and measured for equal accumulations
of proton charge at the spallation target. Measurements were made
with quasi-monochromatic neutron beams at mean wavelengths of 2.41~Å
(14.1~meV), 1.85~Å (24.0 meV), 1.28~Å (49.9 meV) and 0.72~Å (160
meV). Sample-out measurements were also made and subtracted from sample-in
data. Fig.~\ref{fig:Merlin-2D} compares 2D plots of polar scattering
angle vs. energy transfer for the Al sample and sample-out background.
There is considerable background at the forward angles and therefor
the present analysis employed scattering angles $45\lyxmathsym{\textdegree}\leq\theta\leq135\lyxmathsym{\textdegree}$.

\begin{figure}[H]
\begin{center}\includegraphics[width=0.8\columnwidth]{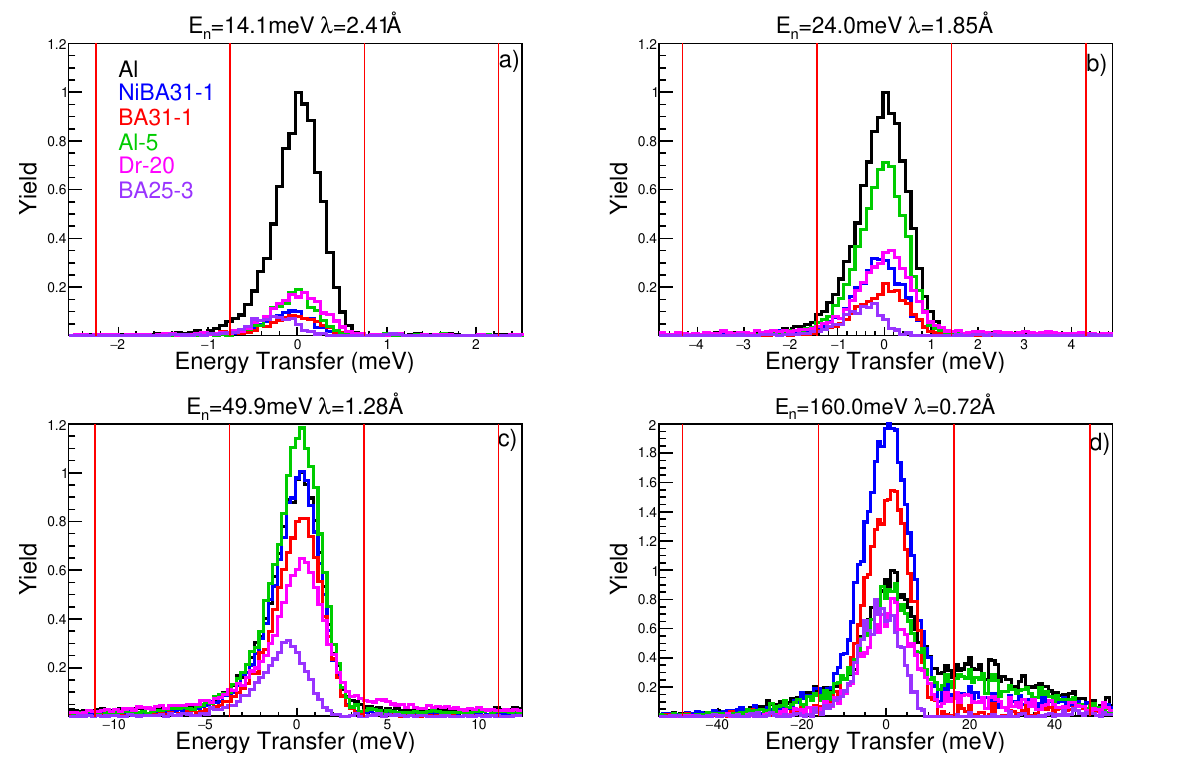}\end{center}

\caption{\protect\label{fig:Energy-transfer-spectra}Energy transfer spectra
summed over scattering angles $45\lyxmathsym{\protect\textdegree}\protect\leq\theta\protect\leq135\lyxmathsym{\protect\textdegree}$.
Sample ID is explained in Table~\ref{tab:Shielding-candidates}.
The red vertical lines denote limits for integration of the elastic
peak and shoulder regions to the left and right of the peak. The width
of each region is $6\sigma$, where $\sigma$ was obtained from Gaussian
fits to the Al peaks at each wavelength.}
\end{figure}

Fig.~\ref{fig:Energy-transfer-spectra} shows the energy-transfer
spectra, summed over angles $45\leq\theta\leq135\lyxmathsym{\textdegree}$,
for the 4 beam energies and 6 scattering samples. Gaussian fits to
the Al distributions produced elastic-peak widths $\sigma.$ The inner
red lines sitting at $\pm3\sigma$ from zero define the elastic scattering
region, while the outer red lines show shoulder regions which range
from $-9\sigma$ to $-3\sigma$ and $+3\sigma$ to $+9\sigma$. 

As Al is the main structural material of MG neutron spectrometers,
the Al data have been employed as a basis with which the other samples
have been compared. Thus the neutron yield scales of Fig.~\ref{fig:Energy-transfer-spectra}
have been normalised so that the peaks of the Al distributions sit
at a value 1.0.

The $\mathrm{^{10}B}$ content in samples is most effective at suppressing
scattering at the longer wavelengths, where the capture cross section
is highest. At 0.72~$\textrm{\AA}$ BA31-1 and NiBA31-1 exhibit more
scattering than Al. They contain $\sim40\%$ more Al than the pure
Al sample and there is an additional contribution to scattering from
the Ni plating on NiBA31-1.

\begin{figure}[H]
\includegraphics[width=1\columnwidth]{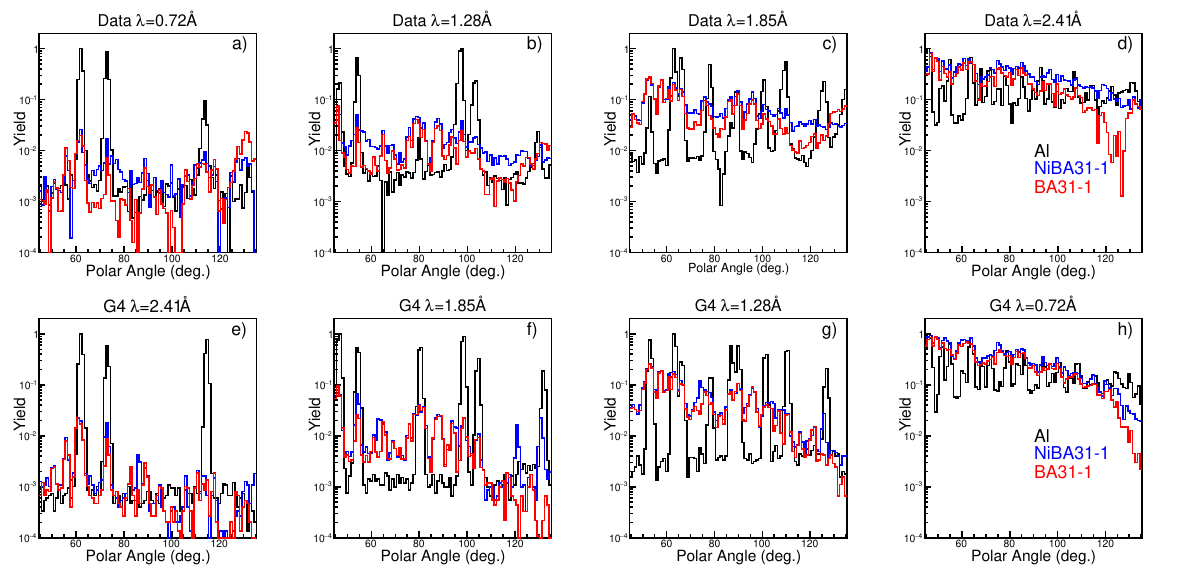}

\caption{\protect\label{fig:Scattering-angle-distributions}Scattering angle
distributions for Al (black), NiBA31-1 (blue) and BA31-1 (red) samples.
Top row: measurements, bottom row: Geant4 simulations. The simulated
distributions have been scaled by a factor $10^{-3}$ to bring them
on to the same range as the measurements.}
\end{figure}

The angular dependence of the scattered neutron yield, integrated
over the elastic peak region, is shown in Fig.~\ref{fig:Scattering-angle-distributions}
(top row) for the Al, NiBA31-1 and BA31-1 samples at the four incident-energy
settings. Similar to Fig.~\ref{fig:Energy-transfer-spectra} the
yields have been normalised so that the peak values of the Al distributions
sit at 1.0.

Prominent Bragg peaks from Al scattering are suppressed in NiBA31-1
and BA31-1, although the composite samples show more incoherent scattering
than pure Al at wavelengths above 2.41~$\textrm{\AA}.$ Comparison
of NiBA31-1 and BA31-1 shows that the Ni plating increases the scattered
yield slightly, but Bragg peaks from Ni are not visible.

Geant4 simulations, which implement a very simple model of the Merlin
setup, are also shown in Fig.~\ref{fig:Scattering-angle-distributions}
(bottom row). Guided by the results of the neutron-transmission measurements
(Table~\ref{tab:Shielding-candidates}), which showed the effective
thickness of $\mathrm{^{10}B}$ in a sample, Merlin simulations were
preformed with a $4\:\mu$m $\mathrm{B_{4}C}$ coating on the Al-5
sample and a $40\:\mu$m dripped coating on the DR-20 sample. Ni thickness
in NiBA31-1 was set to $20\:\mu$m, based on subsequent measurements,
and 9\% P (by weight) incorporated, in line with the plating supplier's
specification. 

The model embodies the polar and out-of-plane angular coverage of
the $\mathrm{^{3}He}$ counter array, but has no fine detail of the
neutron detectors or the materials of the spectrometer in close proximity.
\textcolor{black}{An absolute comparison of the simulated neutron
yields to the data was not possible, due to uncertainties in the data
normalisation procedure. Similar to the data in Fig.~\ref{fig:Scattering-angle-distributions}
the simulated angular distributions have been scaled so that the peak
values for Al sit at 1.0.} The Geant4 model includes the NCrystal
package \citep{ncrystal}, developed at ESS, to model the interaction
of thermal neutrons with Al, Ni, $\mathrm{B_{4}C}$, Al/$\mathrm{B_{4}C}$
composite and epoxy/$\mathrm{B_{4}C}$ dripped coating. It reproduces
the general features of the Bragg-peak structure of Al and the attenuation
of these peaks in the composite samples. A fully realistic model of
Merlin is beyond the scope of this work and detailed differences in
the angular distributions may well be due to the simple model of Merlin
employed. For strongly absorbing samples the backward angle region
close to $135\lyxmathsym{\textdegree}$ is sensitive to the angle
of the sample with respect to the beam (nominally $45\lyxmathsym{\textdegree}$,
Fig.~\ref{fig:Merlin-schematic}). Even a few degrees of misalignment
has a sizeable effect on the backward-angle Geant4 calculation.

\begin{figure}[H]
\includegraphics[width=1\columnwidth]{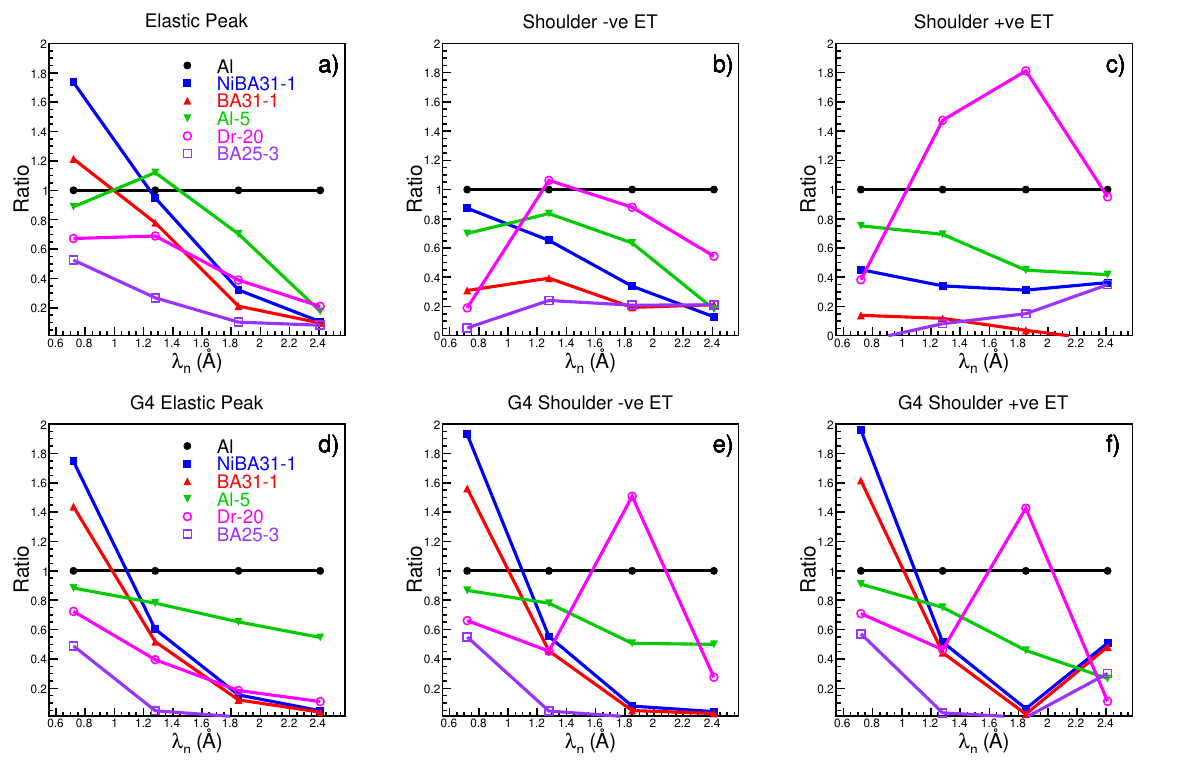}

\caption{\protect\label{fig:E-trans-integrals}The dependence of scattered
neutron yield ratios on scattering sample and incident neutron wavelength
for the elastic and shoulder regions of the energy-transfer spectra
(Fig.~\ref{fig:Energy-transfer-spectra}). The top row shows measured
values, while the bottom row shows simulated values.}
\end{figure}

The energy-transfer spectra of Fig.~\ref{fig:Energy-transfer-spectra}
were integrated over the elastic peak region and the shoulder regions
to the left (-ve ET) and right (+ve ET) of the elastic peak. Fig.~\ref{fig:E-trans-integrals}
compares ratios, the quotients of the sample integral and the Al-sample
integral at a given $\lambda_{n}$. Thus the Al points sit at a value
of 1.0. At longer wavelengths ($\lambda_{n}=1.85,2.41\:\textrm{\AA}$)
the scattered neutron yield in the elastic region is lower, relative
to Al, in samples with $\mathrm{^{10}B}$ content. However at $\lambda_{n}=0.72\:\textrm{\AA}$,
NiBA31-1 and BA31-1 show increased scattering compared to Al. In the
shoulder regions at all measured wavelengths the scattered yield is
lower where the $\mathrm{^{10}B}$ content is higher, apart from Dr-20,
where the base material of the drip coating is epoxy resin. 

As in Fig.~\ref{fig:Scattering-angle-distributions}, the simulation
reproduces the general features of the data in the elastic-peak region.
However the Merlin model, which employs a simple Gaussian TOF structure
for the chopped beam, shows larger discrepancies with the data in
the shoulder regions, notably for the BA31-1 and NiBA31-1 samples
at $\lambda_{n}=0.72\:\textrm{\AA}$.

\section{\protect\label{subsec:Conclusions}Conclusions and Outlook}

Internal shielding is important on Multi-Grid type neutron detectors
to suppress multiple scattering, which can distort measurements of
energy and momentum transfer. $\mathbf{\mathrm{B_{4}C}}$ provides
effective shielding for thermal neutrons, especially if the $\mathrm{^{10}B}$
content is enriched, but can be difficult to incorporate in mechanical
structures. 

Up to now the MG have been made from high purity Al coated with highly
enriched $\mathbf{\mathrm{B_{4}C}}$ and for the normal blades of
a grid a thin sputtered coating works well. The normal-blade coating
must be thin to allow the charged products of $\mathrm{n+^{10}B\rightarrow^{4}He+^{7}Li}$
to escape into the gas of the VPC. For radial blades sputtering can
produce coating thickness up to $\sim5\:\mu$m (sample Al-5) and the
present measurements show significantly lower neutron transmission
and scattering relative to bare Al. Monte Carlo models of T-REX suggest
that use of Al-5 on the radial blades is indeed beneficial, but the
benefit is increased substantially if the $\mathrm{^{10}B}$ thickness
on the radial blades is increased. Transmission measurements on the
present sample of Al-5 show that the average coating thickness is
actually $\sim4\:\mu$m. Merlin measurements show that the coating
reduces the amount of neutron scattering compared to bare Al at wavelengths
greater than 1.28~$\textrm{\AA}$.

Dripped coatings of a $\mathbf{\mathrm{B_{4}C}}$-epoxy mix (Dr-10,
Dr-20) on an Al substrate have lower transmission than Al-5, although
the epoxy component increases the probability of neutron scattering,
especially in the non-elastic region. The present measurements suggest
some difficulty to control the thickness and uniformity of the coating.
Also the long term adhesion of the coating to the Al substrate is
not known.

Cold-spray coating of a $\mathrm{B_{4}C}$/Al composite on to an Al
substrate (Sp1.4 and SP1.6) also has lower transmission than Al-5,
but the $\mathrm{^{10}B}$ content in the two provided samples showed
large differences in the transmission measurement. The coating thickness
shows a banded structure, presumably reflecting the scanning of the
spray gun across the sample. Cold-spray coated samples were not available
for the Merlin scattering measurement. 

Overall, uncoated $\mathrm{B_{4}C}$/Al composite samples performed
best in terms of low neutron transmission and scattering. The transmission
measurements show that the manufacturer's specification of $\mathrm{^{10}B}$
content is quite accurate and the concentration appears fairly uniform
throughout the volume of the sample. It is electrically conductive
and can be water-jet cut and laser welded (if Ni plated). The thinnest
available sheet (BA31-1) is thus suitable for radial blades. However
the Al used in the composite contains traces of actinide impurities
and so the sheet emits alpha particles, which are detected to produce
a constant background counting rate \citep{alpha-background}. A $\sim25\:\mu$m
Ni plating of the composite (NiBA31-1) is highly effective at suppressing
this background and adhesion of the plating to the substrate appears
to be good. However the Ni coating increases neutron scattering relative
to BA31-1. Of the samples tested at Merlin, BA25-3 had the lowest
neutron scattering yield. Thus the thicker composite sheets, which
are also highly opaque to neutrons, will be suitable for additional
shielding at the rear or side of the MG structures.

Based on the measurements reported here, a Multi-Grid prototype for
T-REX, named TRP-3, has been constructed using NiBA31-1 for the radial
blades and BA31 (4~mm thick) for the shielding attached to the rear
of the grid (Fig.~\ref{fig:Multigrid}). The latter reduces back
scattering from Multi-Grid support structures. As NiBA31-1 is an effective
neutron absorber no additional side shielding has been included. TRP-3
has been tested at ESS and ISIS in comparison to an earlier unshielded
prototype TRP-1 and preliminary analyses show that the new internal
shielding is highly effective at suppressing internal neutron scattering.

\section*{Acknowledgements}

We wish to thank the following for their invaluable assistance:
\begin{itemize}
\item the ISIS staff for the efficient provision of the neutron beams at
the EMMA and Merlin facilities, 
\item the technicians of the ESS Detector Group and the University of Glasgow
engaged in the design and construction of Multi-Grid structures,
\item \textcolor{black}{Y. Yang and L. Liu of Tsinghua University, Beijing,
China, who provided the drip coated sample,}
\item G. Zuzel and M. Czubak, of the Jagiellonian University, Kraków, Poland,
who performed alpha background measurements of selected samples.
\end{itemize}
\textcolor{black}{The University of Glasgow acknowledge that the result
has been generated in collaboration with and through financial support
by European Spallation Source ERIC under Contract 325103. Glasgow
also acknowledge support from the UK Science and Technology Facilities
Council, Grant ST/V00106X/1.}

\end{document}